\documentclass[a4paper,10pt,twoside]{cpc-hepnp}

\usepackage{multicol}
\usepackage{graphicx}
\usepackage{booktabs}
\usepackage{amssymb,bm,mathrsfs,bbm,amscd}
\usepackage[tbtags]{amsmath}
\usepackage{lastpage}
\usepackage{CJK}

\begin{document}

\fancyhead[c]{\small Submitted to Chinese Physics C} \fancyfoot[C]{\small 010201-\thepage}

\footnotetext[0]{Received 14 March 2009}

\title{{Study on the novel neutron-to-proton concept for improving the
detection efficiency of triple GEM based fast neutron detector\thanks{Supported by National Natural Science
Foundation of China (11135002, 11305232 and 11175076)}}}

\author{%
 WANG Xiao-Dong$^{1}$
\quad YANG He-Run$^{2}$
\quad REN Zhong-Guo$^{3}$
\quad ZHANG Jun-Wei$^{2}$\\
\quad YANG Lei$^{1}$
\quad ZHANG Chun-Hui$^{1}$
\quad HA Ri-Ba-La$^{4}$
\quad AN Lv-Xing $^{5}$\\
\quad HU Bi-Tao$^{1;1}$\email{hubt@lzu.edu.cn}%
}
\maketitle

\address{%
$^1$ {School of Nuclear Science and Technology, Lanzhou University, Lanzhou 730000, China}\\
$^2$ {Institute of Modern Physics, Chinese Academy of Sciences, Lanzhou, 730000, China}\\
$^3$ {China Academy of Engineering Physics, Mianyang 621907, China}\\
$^4$ {Inner Mongolia Center for Disease Control and Prevention, hohhot, 010031, China}\\
$^5$ {LWD Logging Center, China Petroleum Logging CO.LTD., Xi'an, 710000, China}

}

\begin{abstract}
 A high-efficiency fast neutron detector prototype based on a triple Gas Electron Multiplier (GEM) detector, which coupled with a novel multi-layered High-Density PolyEthylene (HDPE) as a neutron-to-proton converter for improving the neutron detection efficiency, is introduced and tested with the Am-Be neutron source in Institute of Modern Physics (IMP) at Lanzhou in present work. Firstly, the developed triple GEM detector is tested by measuring its effective gain and energy resolution with $^{55}$Fe X-ray source to ensure that it has a good performance. The effective gain and obtained energy resolution is 5.0$\times$10$^{4}$ and around of 19.2\%, respectively. And secondly, the novel multi-layered HDPE converter is coupled with the cathode of the triple GEM detector make it a high-effective fast neutron detector. And its effective neutron response is four times higher than that of the traditional single-layered conversion technique when the converter layer number is 38.
\end{abstract}



\footnotetext[0]{\hspace*{-3mm}\raisebox{0.3ex}{$\scriptstyle\copyright$}2013
Chinese Physical Society and the Institute of High Energy Physics
of the Chinese Academy of Sciences and the Institute
of Modern Physics of the Chinese Academy of Sciences and IOP Publishing Ltd}%

\begin{multicols}{2}

\section{Introduction}
 In recent decades, there was much interest in exploring the use of the neutron-linked techniques for non-destructive screening\cite{lab1}, elemental characterization of bulk materials\cite{lab2} and medical radiotherapy\cite{lab3}. Because neutron is neutral with an unique strongly penetrating power, and cannot be detected by a norm detector. An important issue is how to detect it if it does not directly ionize and cannot be detected by a normal detector. To make a detector sensitive to the neutron, it is necessary to employ a kind of conversion material which allows neutral neutron to produce some charged particles such as proton,$^4$He and $^3$H. Usually the $^{3}$He gas is considered as the best choice to detect the thermal neutron which can be captured by ${^3}$He with  high cross section. Unfortunately, after the terrorist attacks of September 11, 2001, the federal government started to deploy neutron detectors at the U.S. border to help the national security against smuggled nuclear and radiological material\cite{lab4}. As a consequence, the truth of $^{3}$He gas stockpile shrank and horrible lofty price is triggering  interests and stimulating technological efforts in finding out a substitute or an effective platform to replace the $^{3}$He tubes.

 The original motivation of intention of the GEM detector is to detect charged particles, due to its excellent performances\cite{lab5,lab6}, such as withstanding electrical discharge without damage, good position resolution, high effective gain and an adjustable active region and so on. It became a neutron detector by coupling with a conversion layer, which represents a kind of platform in both thermal and fast neutron detection fields\cite{lab7,lab8,lab9,lab10}. Two very common neutron interactions used for detecting the thermal neutron are the reactions of

 n+$^6$Li$\to$$^4$He+$^3$H+4.79 MeV  and  n+$^{10}$B$\to$ $^7$Li$^{*}$+$^4$He
 \begin{center}
  $\to$$^4$He+$^{7}Li$+$\gamma$(0.48MeV)+2.31 MeV(93$\%$)\\
 $\to$$^7$Li+$^4$He+2.79 MeV \qquad \qquad  (7$\%$) \cite{lab11}.
\end{center}

 However, low atomic number materials, such as PolyEthylene (PE), acrylonitrile (C$_{3}$H$_{3}$N), usually have low cost and relatively higher elastic scattering cross sections for detecting the fast neutrons. In addition, H(n,n$^{'}$)p process among the fast neutron interactions with Hydrogen-rich materials is dominant. Regardless of the thermal neutron or the fast neutron, the charged particles emitted as a result of neutron interaction with converters, such as $^{10}$B, $^{6}$Li and Hydrogen, can be easily detected by the triple GEM detector, in which the charged particles create detectable signals by ionizing gas atoms and finally are collected by read-out anode of the detector.\\

 Usually, the detection efficiency of fast neutrons is very low, less than 0.1$\%$\cite{lab12,lab13} and therefore how to improve it is also a very important issue elaborated in this work. The high-efficiency fast neutron detector developed and tested in present work consists of a triple GEM detector and a converter. The triple GEM detector is constructed and tested with $^{55}$Fe source firstly. And then, a single-layered HDPE converter is coupled with the triple GEM detector to make it work as a neutron detector. At last, to improve the neutron detection efficiency, a novel multi-layered HDPE as the neutron-to-proton converter is installed in the triple GEM detector and tested by measuring neutron deposited energy spectrum with Am-Be source as done in the Refs.\cite{lab9,lab14}.

 This paper is organized as follows: in the Section 2, the principal of the high-effective neutron detector based on the triple GEM and experimental setup is introduced; in the Section 3, the triple GEM detector is tested; in the Section 4, the fast neutron detector based on the triple GEM detector is studied and the energy deposition spectrum and relative neutron detection efficiency is presented; the last Section is the conclusions and discussions.

\section{High-effective fast neutron detector and experimental setup description}

The high-effective fast neutron detector is a four-stage parallel plate avalanche chamber and works in a proportional mode as shown in Fig.1. It consists of a cathode plane, three GEMs, and a read-out anode (PCB). The 2 $\mu$m thick Al foil is coated on the multi-layered HDPE converter as the cathode, and each converter layer is 1 mm thick. As shown in Fig.2, two different conversion structures are used for testing in this work, including traditional single layer marked as A and novel multi-layered conversion structures marked B and C, (the layer number of A mode and B mode is 21 and 38, respectively). The detector has a volume of 290$\times$290$\times$56 mm$^{3}$, a square active area of 100$\times$100 mm$^{2}$, a 4 mm wide drift gap, where the process of primary ionization happens, two 2 mm wide transfer gaps, where all electrons transfer toward the induction gap, and a 4 mm wide induction gap, where all avalanche electrons are induce as the detectable signal to the read-out PCB (Printed Circuit Board) anode. The anode plane is made of one-dimensional parallel strips with standard printed circuit board technology. All read-out strips (496) are shorted together for acquiring an integrated anode signal and connected to the electric readout system. And the width of a strip is 80 $\mu$m and the interval between two strips is 110 $\mu$m.
\begin{center}
\includegraphics[width=8.5cm,height=4.5cm]{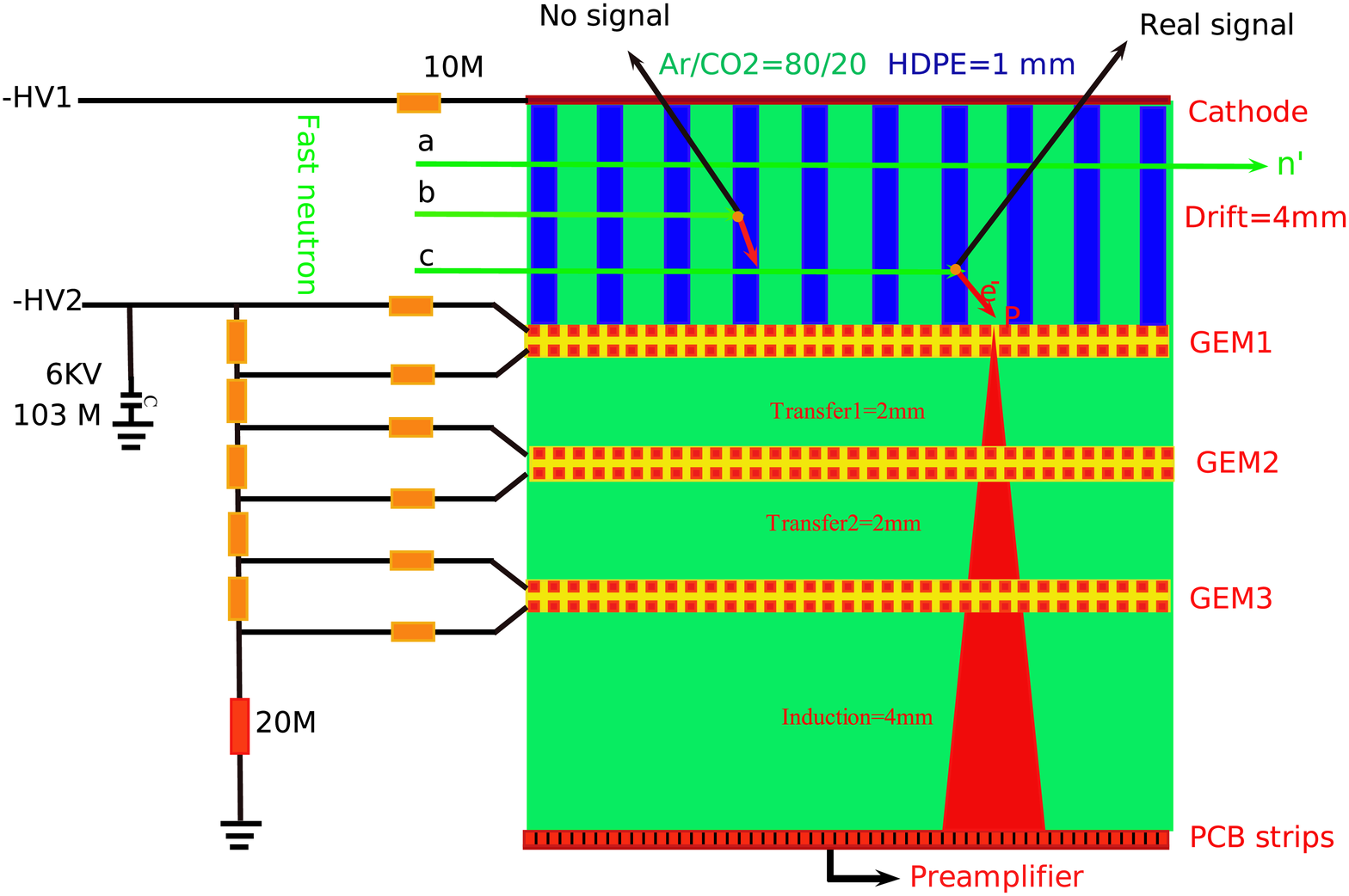}
\end{center}
\figcaption{\label{fig1} (color online) Experimental principal scheme of the high-effective fast neutron detector. The a, b and c represents three different physics processes which are: a) a neutron directly goes through the converter without any interaction; b) a recoiling proton knocked out by a neutron$'$s elastic scattering with the converter yet has not enough energy to escape the converter and terminates in the converter; c)a neutron knocks out a energetic recoiling proton enable to escape the converter, ionize the gas atoms and produce a effective signal in detector. The yellow blocks present the resistors. Other information is described in the scheme.}

The detector is operated based on a continuously flushed Ar/CO$_2$ gas mixture (80/20 percentage in volume). The bias voltage is delivered to each electrode through a passive resistor divider by CAEN module N472. An addition 10 M¦¸ protection resistor connected with GEM electrode is placed in series for releasing discharges. To obtain the pulse height spectrum, the signal is read out by an electronic system comprised of preamplifier Ortec 142PC, amplifier CAEN N968, Ortec CF8000, Ortec GG8000 and Philip ADC 7164 and other CAMAC modules.

The detector is tested with the Am-Be source in Institute of Modern Physics (IMP) of Lanzhou.
\begin{center}
\includegraphics[width=8cm,height=4cm]{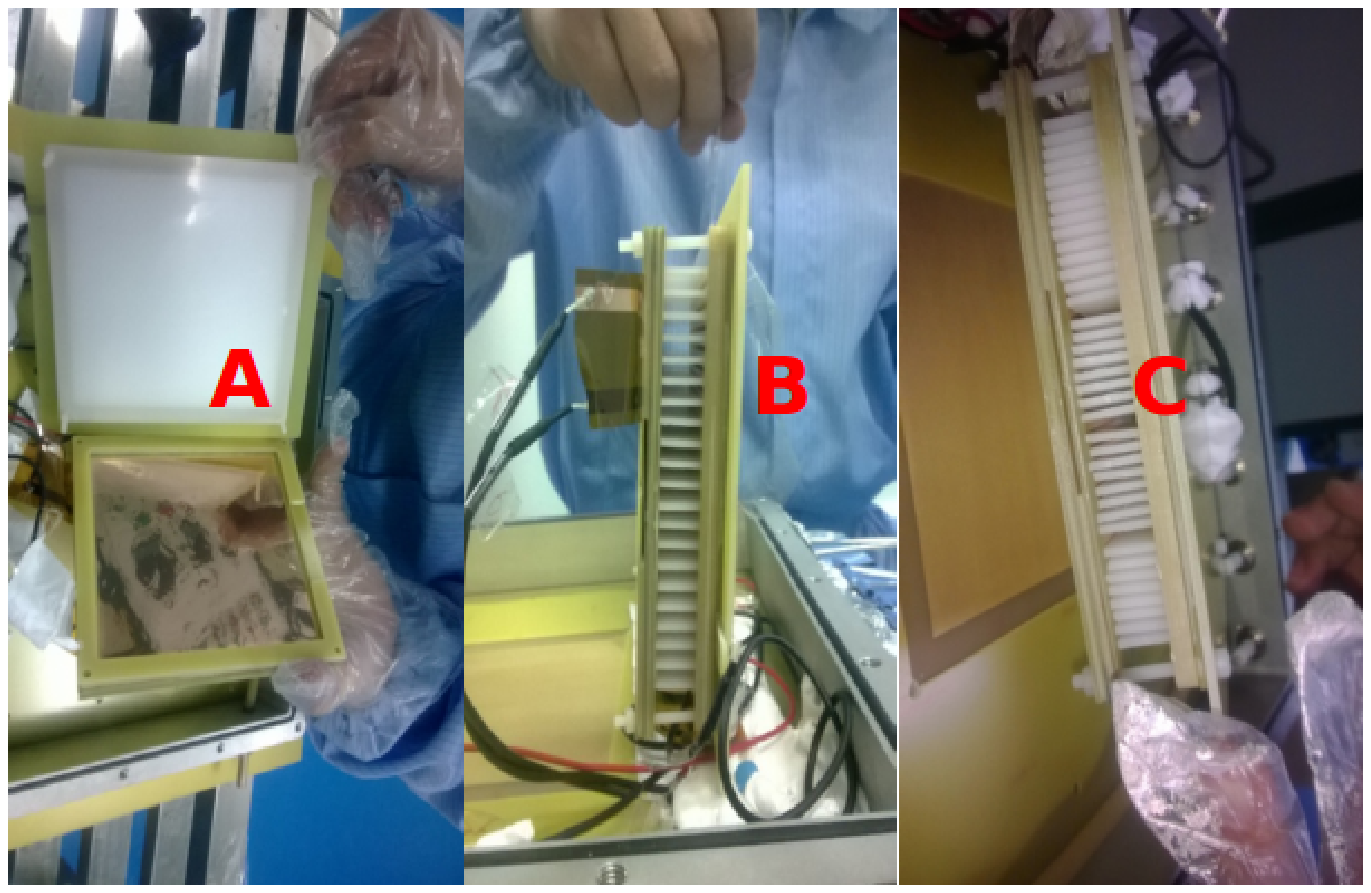}
 \end{center}
 \figcaption{\label{fig2} (color online)Two different conversion structures are employed in present work, including traditional single-layered and novel multi-layered HDPE. A, B and C present the number of layer which is 1, 21 and 38, respectively.}
\begin{center}
\includegraphics[width=8cm,height=6cm]{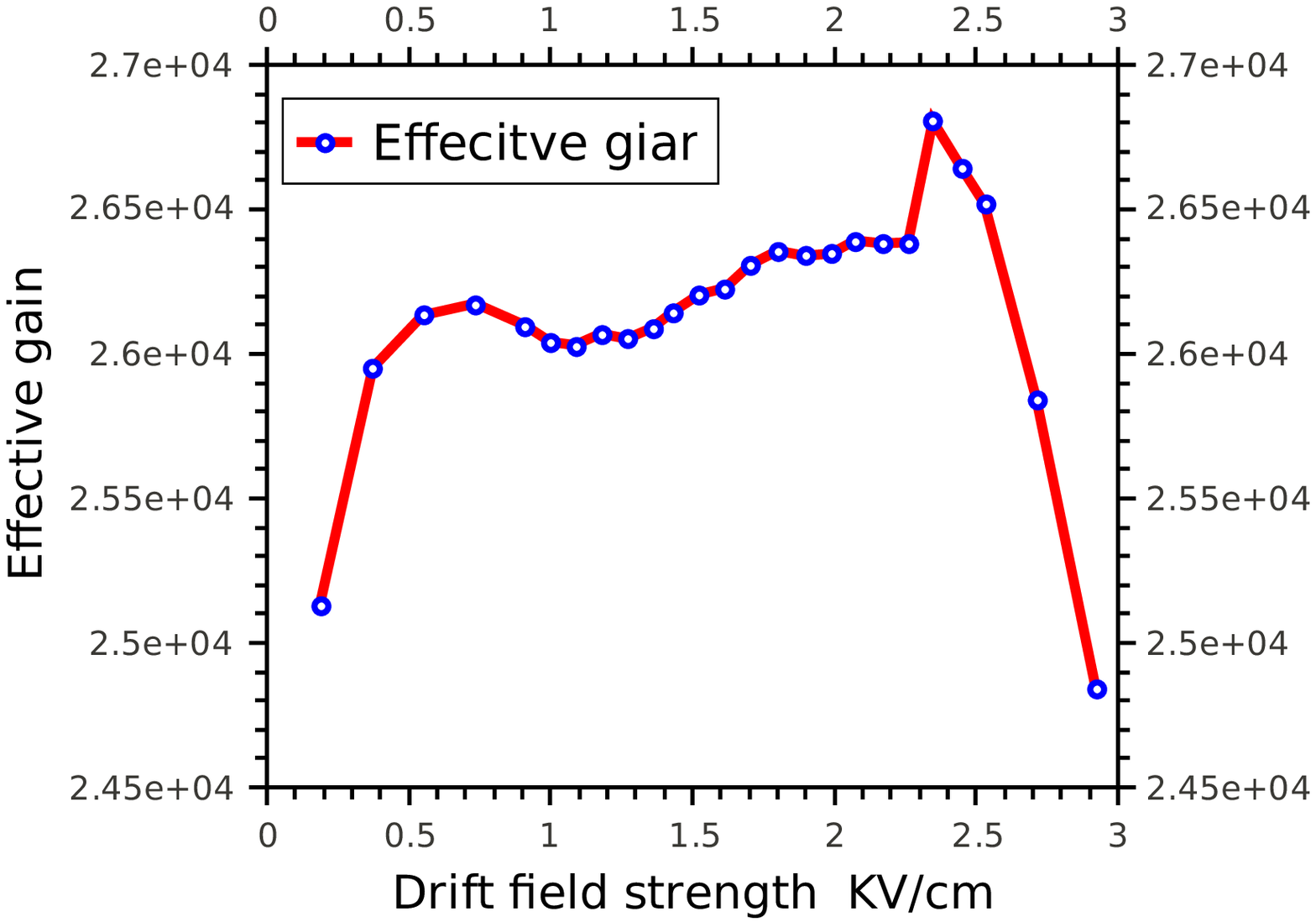}
 \end{center}
 \figcaption{\label{fig3} (color online) Effective gain of the triple GEM detector as a function of E$_{D}$ with $^{55}$Fe 5.9 keV X-ray, V$_{TGEM}$ = 991 V and induction field is 2 KV/cm.}

 \section{The basic performances of triple GEM detector }
\subsection{Effective gain }

Effective gain is a very important character for the GEM detector. Since it impacts, in generally, the energy resolution, the time resolution and position resolution. Three common methods are often employed to improve effective gain in principle. Firstly, by improving the drift field, the ionization electrons can obtain enough kinetic energy to avoid the recombination effect, and go through the GEM hole to produce an avalanche. Secondly, by improving the high voltage across the GEM electrodes, the electronic field inside the GEM hole becomes stronger and the electrons passing by it can obtain enough kinetic energy between two collisions to trigger further secondary ionization. Thirdly, higher induction field can drag the avalanche electrons out of the GEM holes more easily to produce detectable signals. In present work, the effective gain of the triple GEM detector in different drift E$_D$ and the high voltage across the three GEM electrodes V$_{TGEM}$(V$_{gem1}$+V$_{gem2}$+V$_{gem3}$) is studied, respectively. $^{55}$Fe X-ray source is collimated by an iron collimator with a 2 mm hole on the top of the detector.

The effective gain as a function of E$_D$ is shown in Fig.3, which contains three different important components:

i)At the beginning, the effective gain is small. That is because the strength of drift field is too weak to separate the electrons from electron-ion pairs, which results in the  recombination of the electron-ion pairs.

ii)When the drift field gradually becomes stronger, the effective gain has a harshly growing and reaches a saturation plateau. That is because the drift field is strong enough to make the most of ionization electrons free and go through the GEM holes to create an avalanche. Since the total number of electron-ion pairs is limited, with the drift field increasing, the total electrons, which are free of electron-ion pairs and drift through the GEM holes to produce an avalanche, are also limited. That is why the effective gain stays on plateau when the drift field is between 0.5-2.4 KV/cm.

iii)When the drift field is higher than 2.4 KV/cm, the ionization electrons hit the copper of the GEM foil instead of drifting out of the GEM holes, which results in the effective gain aggressively down as shown in Fig.3.

As shown in Fig.4 the effective gain of the detector increases in a good exponential way as the V$_{TGEM}$ creasing. When the V$_{TGEM}$ is up to 1017 V, the effective gain reaches a maximum value of 5.0$\times$10$^4$, where the preamplifier 142 PC appears saturation in this work.

\subsection{The energy resolution}
The energy resolution of the triple GEM detector is tested with $^{55}$Fe 5.9 keV X-ray. Fig.5 shows the energy spectrum under conditions of E$_{D}$ = 2 KV/cm, V$_{TGEM}$ = 1017 V and E$_{I}$ = 2 KV/cm. The detector is operated at the effective gain of 5.0$\times$10$^4$. The energy resolution is about 19.2$\%$. The full photo-electron peak of $^{55}$Fe and the escape peak of Argon atom is distinguished completely. The ratio of two peaks is 1.98, which demonstrates the detection system has a good energy linearity relationship. As show in Fig.6, The energy resolution becomes better with V$_{TGEM}$ increasing. Especially, when the V$_{TGEM}$ is up to 1017 V, the energy resolution reaches the best value of 19.2\% in this test.

\begin{center}
\includegraphics[width=7cm,height=7cm]{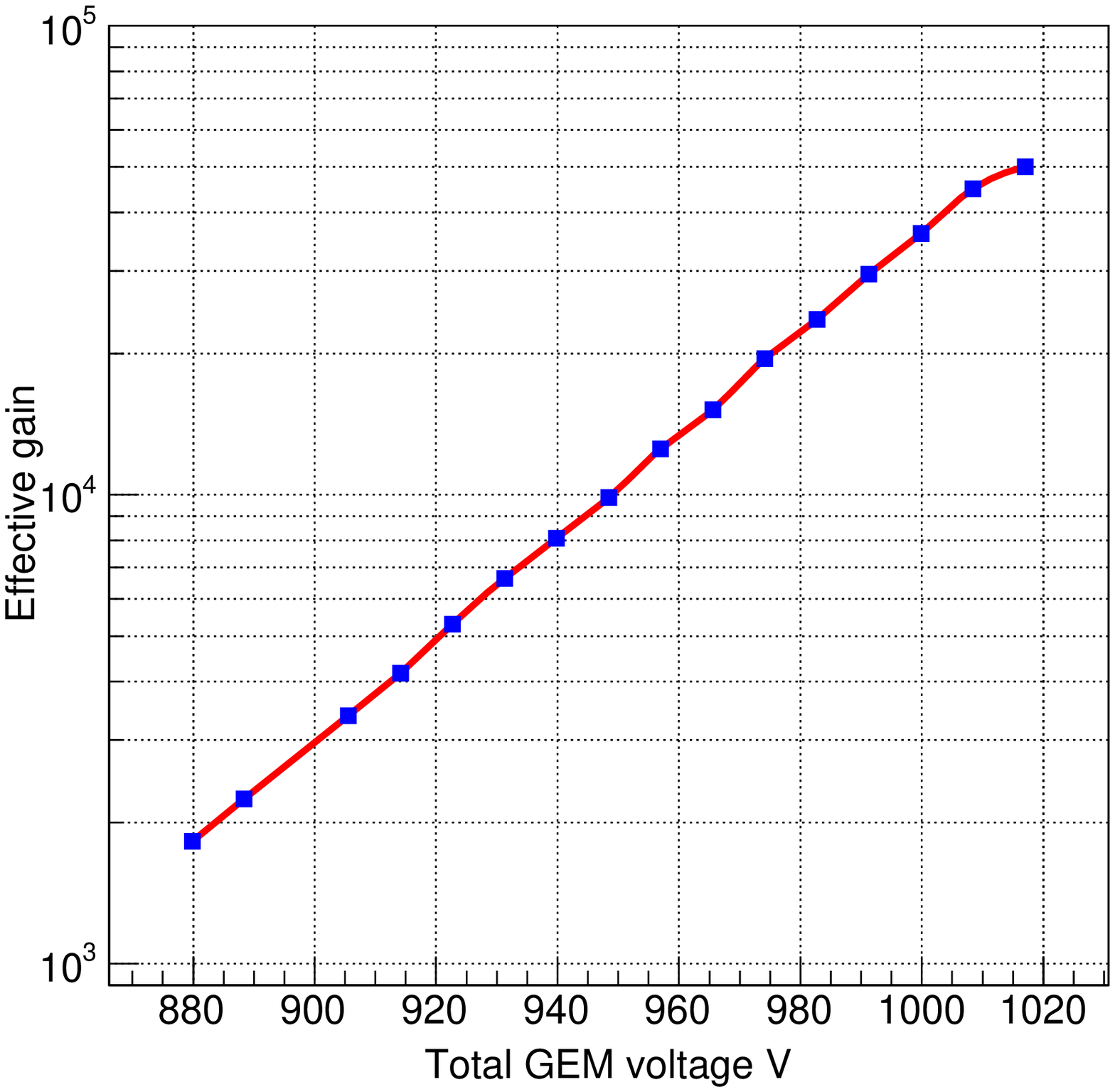}
 \end{center}
   \figcaption{\label{fig4} (color online) Effective gain of the triple GEM detector as a function of V$_{TGEM}$ with $^{55}$Fe 5.9 keV X-ray irradiation under conditions of electric field strength of E$_{D}$ = 2 KV/cm and E$_{I}$ = 2 KV/cm; Ar/CO2=80/20, room temperature.}
\begin{center}
\includegraphics[width=7cm,height=7cm]{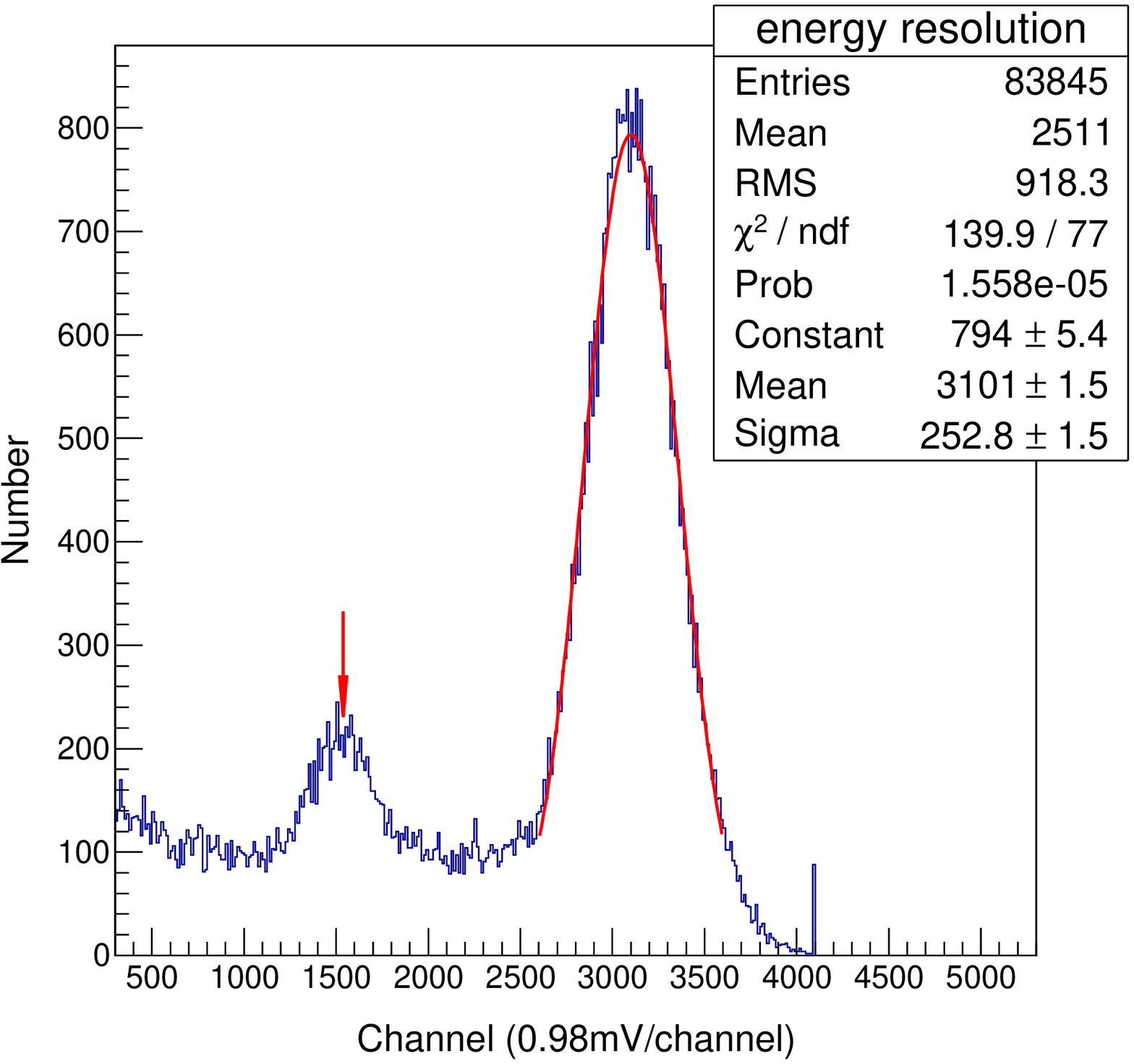}
\end{center}
  \figcaption{\label{fig5} (color online) Energy spectrum of $^{55}$Fe 5.9 keV X ray detected with triple GEM detector with V$_{TGEM}$ = 1017 V, E$_{D}$ = 2 KV/cm, E$_{I}$ = 2 KV/cm, Ar/CO2=80/20, room temperature.}

 \begin{center}
\includegraphics[width=7cm,height=7cm]{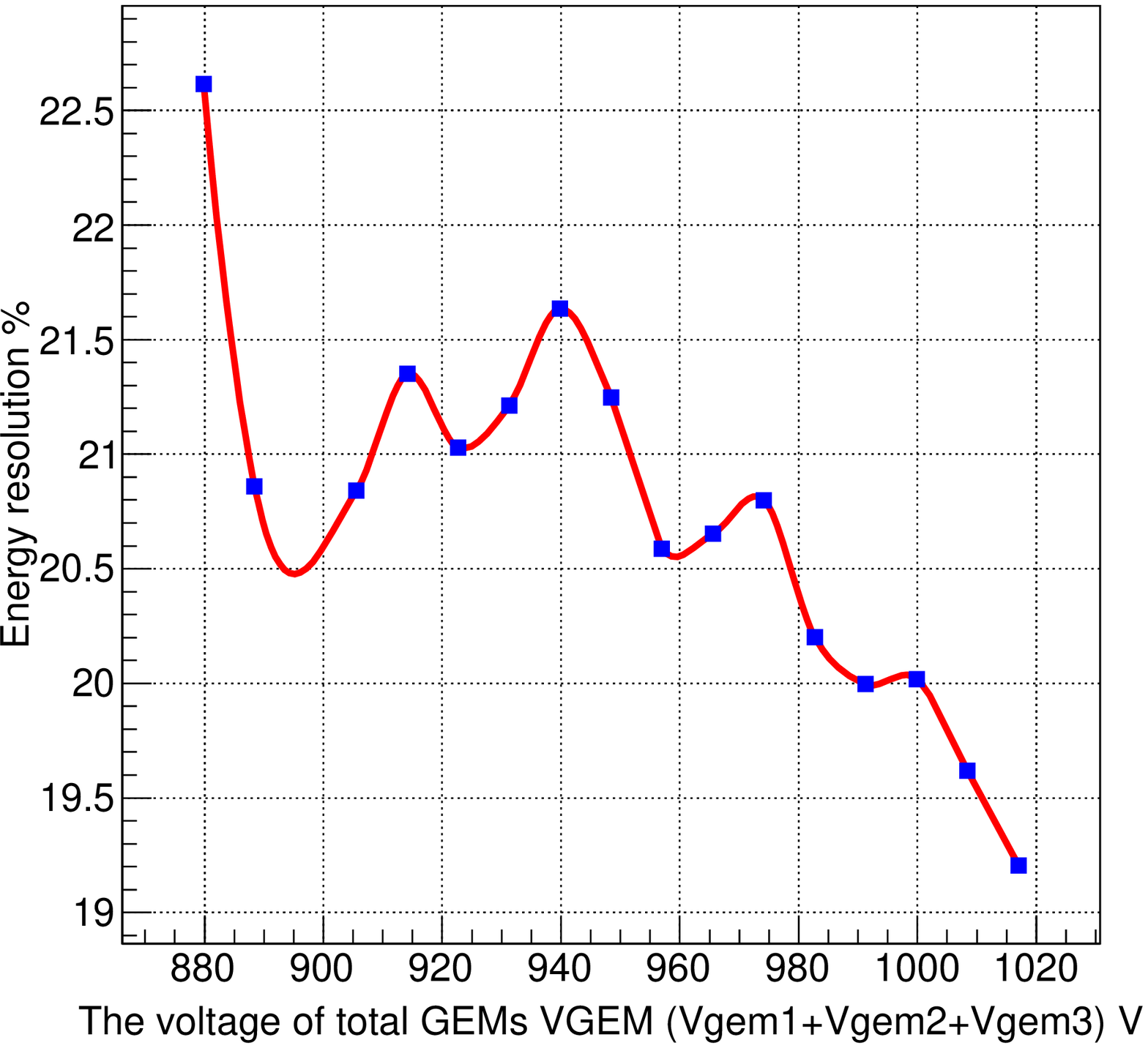}
 \end{center}
 \figcaption{\label{fig6} (color online) Energy resolution of the triple GEM detector as a function of V$_{TGEM}$ with $^{55}$Fe 5.9 keV X-ray, drift and induction field is 2 KV/cm and 2 KV/cm, respectively.}

 \begin{center}
\includegraphics[width=8cm,height=8cm]{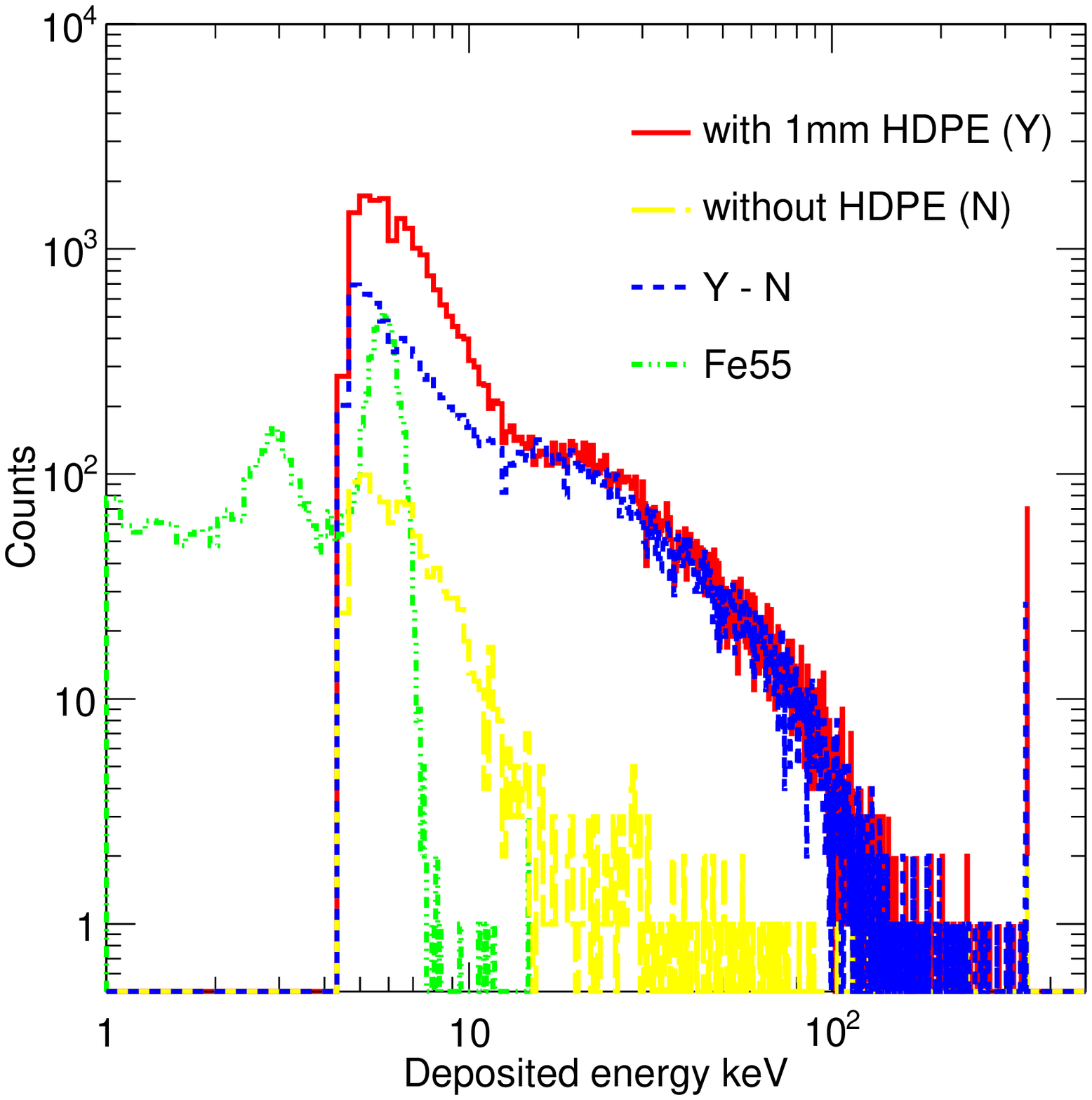}
 \end{center}
 \figcaption{\label{fig7} (color online) Energy spectrum measured with Am-Be neutron. The sample Y (solid line) and N (long dash line) represent the total energy spectrum and background spectrum, respectively. Y and N are obtained with or without the 1 mm HDPE conversion layer under the same other conditions. The sample (Y-N) represents an entire neutron energy deposition. The line of lone dash double points represents $^{55}$Fe X-ray energy spectrum to show the detector working status.}
\section{The results of high-effective fast neutron detector based on triple GEM}

 The single-layered HDPE marked as A in Fig.2 is employed as a traditional technology. In this test, the effective gain of the triple GEM detector is set to 2000 in order to reduce the sensitivity to gamma. The threshold of the CF8000 is set to 21 mV for deceasing the system noise. The neutron response spectrum Y and N were under same conditions except the converter. Y is with 1 mm HDPE, whereas N is without it. The experimental results are shown in Fig.7. In the spectrum (Y-N), three different components can be distinguished clearly. The Refs.\cite{lab15,lab16} found the similar results in simulation and experiment.

  a)The lower than 18 keV part is considered as from the activation photons, which librate electrons by Compton Scattering to deposit their energies in the detector and contribute to the lower energy part of the neutron spectrum.

  b)The higher than 18 keV part is regarded as the real neutron response signals. The charged particles coming from neutron converter of the HDPE and the GEM Kapton deposit higher energy than that the Compton electrons do. This part is called the neutron conversion region in the energy spectrum, which demonstrates the real response events of the detector to neutrons.

  c)The peak around 320 keV is due to the saturation of the employed 142 PC preamplifier.

 As a comparison with the single-layered traditional conversion method, the novel multi-layered HDPE converter is tested, too. The layer number of converter is designed to be 21 and 38 respectively and marked as B and C in Fig.2. The multi-layered conversion detector is measured with the same conditions as that of single-layered. The value of real neutron response event rate is subtracted from the energy spectrum for comparing with the traditional conversion technique. As shown in Fig.8, the results show that the relative detection efficiency of novel multi-layered neutron detector is about four times higher than that of traditional method when the converter layer number is 38.
\section{Conclusions and discussions}

Basic performances of the triple GEM detector are studied by measuring effective gain and energy resolution with $^{55}$Fe X-ray source. The experiment results confirm that the triple GEM detector has a high effective gain about 5.0$\times$10$^{4}$ and a good energy resolution around of 19.2$\%$ under a lower safe high voltage. The neutron GEM-based detectors with single-layered and multi-layered HDPE converters are studied. By comparing with the count of the real neutron response event rate, multi-layered converter technique is higher than that of the traditional signal-layered one. In other words, the multi-layered neutron conversion technique can really greatly improve neutron detection efficiency in practical applications, which opens a new method to detect the neutron in the future.

\begin{center}
\includegraphics[width=8cm,height=8cm]{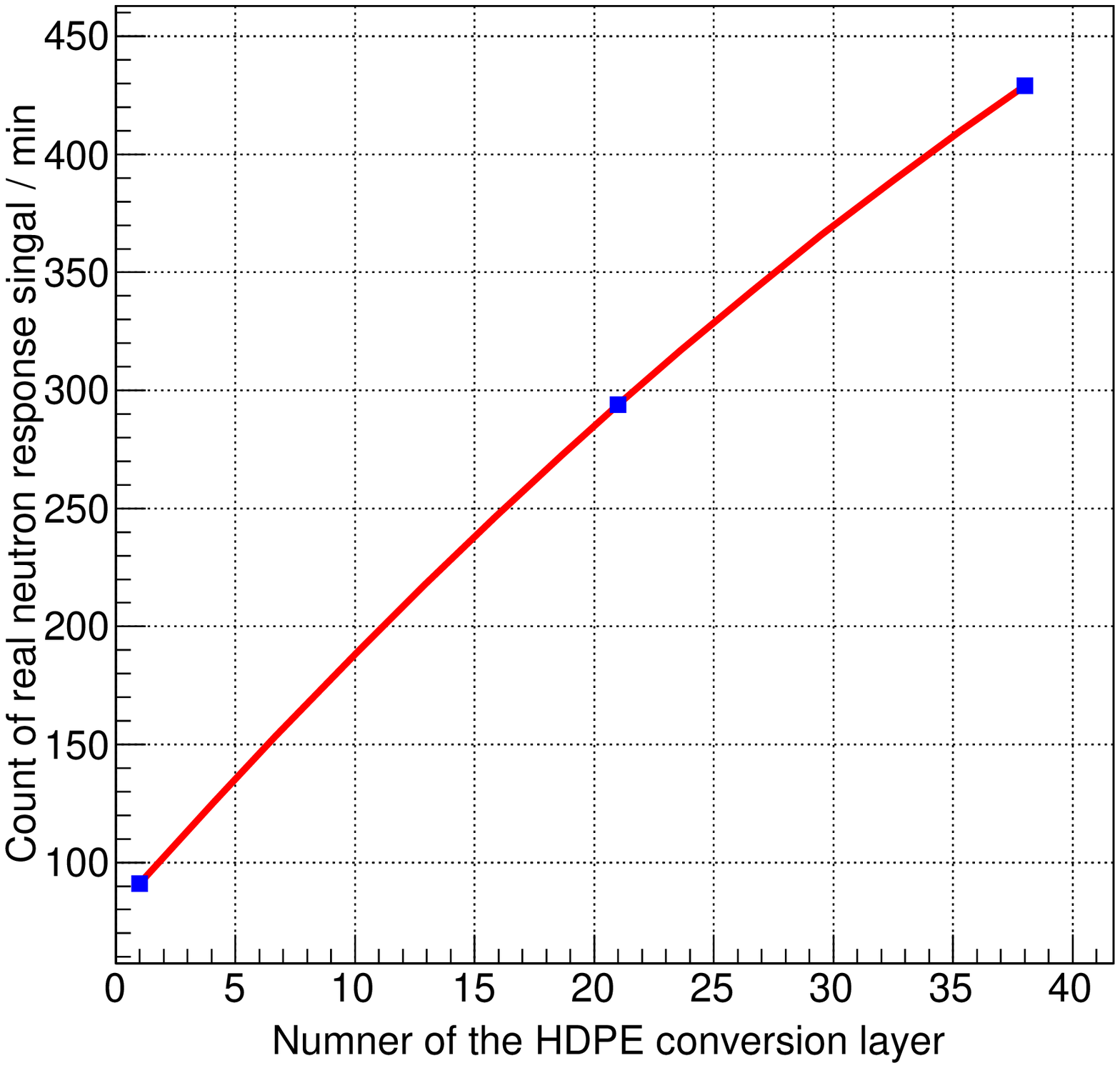}
 \end{center}
 \figcaption{\label{fig8} (color online) The count of real neutron response events as a function of the number of converter layer, the number of the converter layer is 1, 21 and 38 layers, respectively.}

For novel multi-layered conversion technique, due to limitations of the current production technique and mechanical constraints, it is hard to ensure converter surface without burr, which results in HDPE charging up and further lead to distortion of the electric field in drift volume. This affects the efficiency in collecting and focusing the ionization electrons to the GEM holes and thus significantly decrease the total effective events. In addition, because the neutron source used in this test is not a mono-energy such as D-T neutron source, only the relative detection efficiency is presented rather than the absolute detection efficiency.

As discussed above, there are two important things should be done in the next work. They are that the multi-layered converter should be manufactured by 3D print technology to improve the surface quality of the converter, and mono-energy neutron source should be employed to finally figure out the absolute detection efficiency. \\

\acknowledgments{We are grateful to DUAN Li-Min, LU Chen-Gui and HU Rong-Jiang for their helpful discussions and support in Institute of Modern Physics.}

\end{multicols}

\vspace{10mm}

\begin{multicols}{2}

\end{multicols}

\vspace{-1mm}
\centerline{\rule{80mm}{0.1pt}}
\vspace{2mm}

\begin{multicols}{2}

\end{multicols}

\clearpage

\end{document}